
\def\IR{{\hbox{{\rm I}\kern-.2em\hbox{\rm R}}}}
\def\IB{{\hbox{{\rm I}\kern-.2em\hbox{\rm B}}}}
\def\IN{{\hbox{{\rm I}\kern-.2em\hbox{\rm N}}}}
\def\IC{{\ \hbox{{\rm I}\kern-.6em\hbox{\bf C}}}}

\def\IZ{{\hbox{{\rm Z}\kern-.4em\hbox{\rm Z}}}}
\def\to{\rightarrow}
\def\d{{\rm d}}
\def\underarrow#1{\vbox{\ialign{##\crcr$\hfil\displaystyle
{#1}\hfil$\crcr\noalign{\kern1pt
\nointerlineskip}$\longrightarrow$\crcr}}}
%
\def\d{{\rm d}}
\def\ltorder{\mathrel{\raise.3ex\hbox{$<$}\mkern-14mu
             \lower0.6ex\hbox{$\sim$}}}
\def\lesssim{\mathrel{\raise.3ex\hbox{$<$}\mkern-14mu
             \lower0.6ex\hbox{$\sim$}}}

\input phyzzx
\overfullrule=0pt
\tolerance=5000
\overfullrule=0pt
\twelvepoint

\twelvepoint
\pubnum{IASSNS-HEP-92/63}
\date{October, 1992}
\titlepage
\title{SOME COMPUTATIONS IN BACKGROUND INDEPENDENT \break
OFF-SHELL STRING THEORY}
\vglue-.25in
\author{Edward Witten
\foot{Research supported in part by NSF Grant
PHY91-06210.}}
\medskip
\address{School of Natural Sciences
\break Institute for Advanced Study
\break Olden Lane
\break Princeton, NJ 08540}
\bigskip
\abstract{Recently, background independent open-string field theory
has been formally defined in the space of all two-dimensional
world-sheet theories.  In this paper, to make the construction more
concrete, I compute the action for an off-shell tachyon field of a certain
simple type.  From the computation it emerges
that, although the string field action does not coincide with the
world-sheet (matter) partition function in general, these functions do coincide
on shell.  This can be demonstrated in general, as long as matter and
ghosts are decoupled.}
\endpage

\REF\witten{E. Witten, ``On Background Independent Open-String
Field Theory,'' IASSNS/HEP-92/53, to appear in Physical Review D.}

\chapter{Introduction}

The beauty of the world-sheet approach to string perturbation theory
has led many physicists to hope that
``the space of all two-dimensional world-sheet field
theories'' might be a natural arena for string field theory.
This program has been obstructed by (i) the unrenormalizability of the
generic world-sheet theory, and other more or less related problems;
(ii) the fact that even given a ``space of all two-dimensional theories,''
one has not known how to formulate a gauge invariant Lagrangian in
that space.  In a recent paper [\witten], the second problem has been
resolved in the case of open strings.  Dealing with open strings
means that we consider two dimensional Lagrangians of the form
$$ L =L_0 + L', \eqn\arfo$$
where $L_0$ is a bulk action describing a closed string background
and $L'$ is a boundary term describing the coupling
to external open strings.  $L_0$ is kept fixed and $L'$ is permitted to
vary.  For instance, in this paper we will take $L_0$ to describe the
standard closed string background
$$L_0=\int_\Sigma \d^2\sigma\left({1\over 8\pi}h^{\alpha\beta}\partial_\alpha
X^i\partial_\beta X^j\eta_{ij} +{1\over 2\pi}b^{ij}D_ic_j\right), \eqn\barfo$$
while $L'$ will be
an arbitrary ghost number conserving boundary interaction
$$ L'=\int_{\partial\Sigma}\d\theta \,\,\,{\cal V}(X,b,c). \eqn\carfo$$
(Notation is standard: $\Sigma$ is a Riemann surface with metric $h$,
inducing a length element $\d\theta$ on the boundary; $b$ and $c$
are the usual anti-ghosts and ghosts; the $X^i$ describe a map to
26 dimensional Minkowski space with Lorentz metric $\eta$.)
Actually, in the construction in [\witten] one must introduce
a local operator ${\cal O}$ of ghost number 1 with
$${\cal V}=b_{-1}{\cal O}. \eqn\muco$$
For the definition of $b_{-1}$ see [\witten].

The main result of [\witten] was to describe, given this data, a background
independent, gauge
invariant Lagrangian $S$ on the space of ${\cal O}$'s.
The construction has many properties that agree with expectations
from world-sheet perturbation theory but have previously been hard
to understand in the context of a gauge invariant Lagrangian.
For instance, the classical equations of motion derived from $S$
are equivalent to BRST invariance
of the world-sheet theory.  This is an improved version, which does
not assume decoupling of matter and ghosts, of the naive expectation that
the equations of motion should assert conformal invariance.  (If it is the
case that on-shell
one can always decouple matter and ghosts by a gauge condition
on ${\cal O}$, then in that gauge the equations of motion
derived from $S$ are equivalent to conformal invariance.)  Moreover, in
expanding around a classical solution, the infinitesimal gauge
transformations are generated by the world-sheet BRST operator.

The construction in [\witten] was, however,
purely formal, since it involved correlation functions in the theory
\arfo\ for arbitrary ${\cal V}$, and one certainly must expect
ultraviolet divergences.  Here is where one faces the fact that we do not
know what ``the space of two-dimensional field theories'' is supposed
to be.   Nevertheless, the definition of $S$ has the
property, noted at the end of [\witten],
that given any concrete family of two dimensional field theories, the
$S$ function can be computed explicitly as a function on the parameter
space of that family.  The original goal of the present paper was
simply to make the ideas of [\witten]
more concrete by computing the $S$ function
explicitly for a certain simple family of boundary interactions.
This will be accomplished in \S2.

\REF\fradkin{E. Fradkin and A. A. Tseytlin, ``Non-Linear Electrodynamics
{}From Quantized Strings,'' Phys. Lett. {\bf B163} (1985) 123.}
\REF\nappi{A. Abouelsaood, C. G. Callan, C. R. Nappi, and S. A. Yost,
``Open Strings In Background Gauge Fields,''
Nucl. Phys. {\bf B280} (1987) 599.}
{}From the computation will emerge a simple relation between the $S$
function and the partition function of the matter system.
We will explore this relation in \S3.  The main conclusion is as follows:
if matter and ghosts are decoupled, then {\it on shell}, the $S$
function is equal to the partition function $Z$ of the matter system.
This relation has been heuristically expected (and checked in some special
cases [\fradkin,\nappi])
given the role of world-sheet path integrals in generating
effective string interactions.

\chapter{Some Simple Boundary Interactions}
\section{The Model}

The open string tachyon corresponds to a boundary interaction of the form
$${\cal V}=T(X). \eqn\unco$$
For ${\cal V}$ of this form, or more generally any ${\cal V}$ that
depends on $X$ only, not on $b$ and $c$, it is natural to choose
$${\cal O}=c{\cal V}, \eqn\huco$$
with $c$ the component of the ghost field tangent to the boundary.
This is the most general situation that we will consider in this paper.

In the present section, to obtain a family of boundary interactions
for which everything can be computed explicitly, we will take
$T(X)$ to be a quadratic function of the coordinates
$$T(X)={a\over 2\pi}+\sum_{i=1}^{26}{u_i\over 8\pi}X_i{}^2, \eqn\cuco$$
with parameters $a$ and $u_i$.
The quadratic nature of the boundary interaction ensures that the world-sheet
theory is soluble, so that we will be able to evaluate explicitly
all the correlation functions that enter in the definition of
the background independent action $S$.

Before entering into actual calculations, let us make a few general
comments about this family of boundary interactions:

(1) Linear terms in the $X_i$ have been omitted, because as long
as the $u_i$ are non-zero, linear terms can be absorbed in a shift
in the $X_i$.

(2) Once the $u_i$ are included, it would be unnatural not to include
the constant term $a/2\pi$ in the action. This is because under a change
in the normal-ordering prescription used in defining the quantum
operator
$\sum_{i=1}^{26}u_iX_i{}^2$, this operator will be shifted by a constant.
With the tachyon interaction \cuco\ is associated a natural
27 parameter family of quantum theories, but the parametrization of the
family by 26 $u$'s and one $a$ is not completely natural, having the
ambiguity just cited.  The background independent action $S$ is well-defined
as a function on the 27 dimensional space of quantum theories of this
type; identifying it as a function
of $a$ and the $u$'s requires a specific normal-ordering recipe.

(3)  The world-sheet action is bounded below only if the $u_i$ are
positive.  We should therefore expect the space-time action $S$ to
have singularities for negative $u_i$.

(4) For $u_i=0$, the theory is invariant under translations of the $X_i$.
The strings propagate in an infinite volume, and the natural object
to calculate is the action per unit volume.  Taking $u_i>0$ gives
a potential energy for the zero mode of the string, so oscillations
of the string are limited to a finite volume.
If the $u_i$ are positive
and small, the action is $\lesssim 1$ for $X_i\lesssim 1/\sqrt{u_i}$.
One therefore should expect that
$$ S\sim {w\over \prod_i\sqrt{u_i}} ~~{\rm for}~~ u_i\to 0, \eqn\duco$$
where $w$ is a constant proportional to
the action per unit volume at $u_i= 0$.

(5) The theory with $a=u_i=0$ is conformally invariant; this case
corresponds to free boundary conditions, $n^\alpha\partial_\alpha X^i = 0$
on $\partial\Sigma$ ($n^\alpha$  is the normal vector to the boundary
of $\Sigma$).  There is also a conformally invariant theory at
$u_i=\infty$ (with boundary conditions $X=0$ on $\partial\Sigma$).
It would be nice to be able to compare the values of the action of two
different classical solutions, but the present example is not quite suitable
for this,
because the theory at $u_i=\infty$ has an action,
while the theory at $u_i=0$ has an action per unit volume, as explained
in the last paragraph.

In \S2.2 below, we will evaluate basic properties of the model, determining
the essential Green's functions and correlation functions and the partition
function.  Then in \S2.3, we will compute the action function $S$ for this
family of quantum field theories.

\section{First Properties}

Let us focus on a single scalar field $X$ described by the
Lagrangian
$$L={1\over 8\pi}\int_\Sigma \d^2\sigma \sqrt h h^{\alpha\beta}
\partial_\alpha X\partial_\beta X +{u\over 8\pi}\int_{\partial\Sigma}
\d\theta \,\,\,\, X^2. \eqn\ingo$$
The boundary condition derived by varying this action is
$$n^\alpha\partial_\alpha X+uX=0 ~ {\rm on }~\partial \Sigma. \eqn\ucco$$

In keeping with the formulation of [\witten], we wish to consider this
theory on a disc $\Sigma$ with a rotationally invariant metric.
We may as well take this to be the flat metric
$$\d s^2=\d\sigma_1^2+\d\sigma_2^2, ~~~~~ \sigma_1^2+\sigma_2^2
\leq 1.\eqn\nucco$$
We also set $z=\sigma_1+i\sigma_2$.

The Green's function of the theory should obey
$$-{1\over 2\pi}\partial_z\partial_{\overline z}G(z,w)=\delta^2(z,w),
\eqn\pucco$$
along with the boundary condition \ucco.  These requirements determine
the Green's function to be
$$G(z,w)=-\ln|z-w|^2-\ln|1-z\overline w|^2+{2\over u}
-2u \sum_{k=1}^\infty{1\over k(k+u)}\left((z\overline w)^k+
(\overline z w)^k\right).          \eqn\hucco$$
The divergence at $u=0$ reflects the fact that there is a zero mode
(the constant mode of $X$) for $u=0$; the physics behind this was discussed
as point (4) at the end of \S2.1.

We now need to define the quantum operator $X^2(z)$,
for $z\in\partial \Sigma$.
As $\partial \Sigma$ corresponds to $|z|=1$, we can write $z=e^{i\theta}$,
and we write $X^2(\theta)$ for $X^2(z)$.
We define
$$X^2(\theta)=\lim_{\epsilon\to 0}\left(X(\theta)X(\theta+\epsilon)
-f(\epsilon)\right)
        \eqn\mcop$$
where $f(\epsilon)$ is a function of $\epsilon$, but not $u$, chosen
so that the limit exists.  These conditions are obeyed by
$$ f(\epsilon)= -2\ln|1-e^{i\epsilon}|^2.  \eqn\burio$$
$f$ is uniquely determined up to an additive constant
(which is the normal ordering constant mentioned in point (2) at the
end of the last subsection) plus terms that vanish for $\epsilon\to 0$.
The requirement that $f$ is independent of $u$ ensures that when
we compute $u$ derivatives (to evaluate the partition function and the
$S$ function) we need not worry about terms coming from the $u$ dependence
of the definition of $X^2(\theta)$.

With the above choice of $f$, the expectation value of $X^2(\theta)$ is
$$ \langle X^2(\theta)\rangle={2\over u}-4u\sum_{k=1}^\infty {1\over k(k+u)}.
\eqn\cucco$$
(In the present subsection only, the symbol $\langle \dots\rangle$ refers
to normalized correlation functions.)

Now we can calculate the partition function on the disc of the theory
of one scalar field
with Lagrangian \ingo.  We will call this partition function $Z_1(u)$.  From
$$Z_1=\int DX\,\,\,\exp(-L), \eqn\jucco$$
and the explicit form of $L$, we have
$${\d\over \d u}\ln Z_1= -{1\over 8\pi}\int_0^{2\pi}\d\theta \langle
X^2(\theta)
\rangle=-{1\over 2u}+\sum_{k=1}^\infty {u\over k(k+u)}.\eqn\didd$$

\REF\ahlfors{L. Ahlfors, {\it Complex Analysis} (McGraw-Hill, 1979).}
The Euler gamma function $\Gamma(u)$ obeys [\ahlfors,  pp. 198-200]
$${\d\over \d u}\ln \Gamma =-{1\over u}+\sum_{k=1}^\infty {u\over k(k+u)}
-\gamma, \eqn\hobo$$
with $\gamma$ being Euler's constant, so
$${\d\over \d u}\ln Z_1={\d\over \d u}\ln \Gamma +\gamma +{1\over 2u}.
              \eqn\nobo$$
Hence (up to an arbitrary multiplicative constant, which can
be absorbed by adding a constant to
the $a$ parameter in the boundary interaction)
$$Z_1(u)=\sqrt u\cdot \exp(\gamma u)\cdot \Gamma(u). \eqn\xobo$$
Note the expected small $u$ behavior $Z_1(u)\sim 1/\sqrt u$ for $u\to 0$ (where
$\Gamma$ has a simple pole),
and the anticipated singularities
for negative $u$ (from the other poles of $\Gamma$).

For a slightly more general model with several scalar fields $X_i$ and
boundary interaction \cuco, the partition function on the disc is therefore
$$Z(u_i;a)=e^{-a}\cdot \prod_iZ_1(u_i). \eqn\jco$$
The simple $a$ dependence comes from the fact that the $a$ term is just
an additive constant in the Lagrangian.

\subsection{An Identity}

Let us now pause to evaluate a correlation function that is needed later.
First of all, for boundary points $z=e^{i\theta}$, $w=e^{i\theta'}$
the propagator (which we will write as $G(\theta,\theta')$) is
$$G(\theta,\theta')=-2\ln(1-e^{i(\theta-\theta')})-
2\ln(1-e^{-i(\theta-\theta')}) +{2\over u}
-2u\sum_{k=1}^\infty {1\over k(k+u)}\left(e^{ik(\theta-\theta')}
+e^{-ik(\theta-\theta')}
\right). \eqn\omigo$$
Expanding $\ln(1-e^{\pm i (\theta-\theta')})$ in a power series
and collecting terms, one finds
$$G(\theta,\theta')=
2\sum_{k\in \IZ}{1\over |k|+u}\exp\left(ik(\theta-\theta')\right).\eqn\migo$$

Now we want to evaluate
$$W=\int_0^{2\pi}{\d\theta\,\,\d\theta'\over (2\pi)^2}\cos(\theta-\theta')
\langle X^2(\theta) X^2(\theta')\rangle.  \eqn\jimox$$
Using the explicit form of the Green's function and doing the angular
integrals, this becomes
$$\eqalign{W=& 4\int_0^{2\pi}{\d\theta\over 2\pi}(e^{i\theta}+e^{-i\theta})
\sum_{k\in \IZ}{e^{ik\theta}\over |k|+u}
\sum_{k'\in \IZ}{e^{ik'\theta}\over |k'|+u} \cr
= &  4\sum_{k\in\IZ}{1\over |k|+u}\left({1\over |k+1|+u}+{1\over |1-k|+u}
\right)= 8 \sum_{k\in\IZ}{1\over |k|+u}{1\over |k+1|+u}
\cr = & 16\sum_{k\geq 0}{1\over |k|+u}{1\over |k+1|+u} =16\sum_{k \geq 0}
\left({1\over k+u}-{1\over k+1+u}\right) = {16\over u}.\cr}\eqn\longeq$$
This is the desired identity.

\section{Evaluation Of The Action}

The definition of the background independent action $S$ in [\witten] was as
follows.  Suppose that the boundary interaction is ${\cal V}=b_{-1}{\cal O}$
where ${\cal O}$ has ghost number 1.  ($S$ depends on ${\cal O}$, not just
on ${\cal V}$.)  Let $O_i$ be a basis of ghost number 1 operators, so
that ${\cal O}$ has an expansion ${\cal O}=\sum_i w^iO_i$.  Then $S$
is defined\foot{With a different normalization from that in [\witten].}
in terms of correlation functions on the disc by
$${\partial S\over \partial w^i}={1\over 2}\int_0^{2\pi}{\d\theta\,\,\d\theta'}
\left\langle O_i(\theta)\,\,\, \{Q,{\cal O}\}(\theta')\right\rangle,
\eqn\kko$$
or equivalently
$$\eqalign{\d S &={1\over 2}\sum_i \d w^i\int_0^{2\pi}{\d\theta\,\,\d\theta'}
\left\langle O_i(\theta) \,\,\{Q,{\cal O}\}(\theta')\right\rangle\cr & =
{1\over 2}\int_0^{2\pi}{\d\theta\,\,\d\theta'}
\left\langle \d {\cal O}(\theta)\,\,\{Q,{\cal O}\}(\theta')\right\rangle.\cr}
 \eqn\kulko$$
Here and henceforth the symbol $\langle \dots\rangle$ refers to
unnormalized correlation functions (one is not to divide by the partition
function).

Obviously, \kulko\ determines $S$ up to an additive constant.
The fact that $ S$  exists is, however, non-trivial.  It depends on the
fact that the one-form on the right hand side of \kulko\ is closed.
In the example we are considering, the proof of this depends on the
non-trivial formula \longeq.  The general proof depends on
Ward identities discussed in [\witten].

In our case, we have
$${\cal V}={a\over 2\pi} +{1\over 8\pi}\sum_iu_iX_i^2,\eqn\occo$$
and ${\cal O}=c{\cal V}$.  In general, for any tachyon field $T(X)$, one has
$$\{Q,cT(X)\}=c c' \cdot\left(2\sum_i{\partial^2\over\partial X_i^2}
+1\right)T(X)     \eqn\homigo$$
(with $c'$ the tangential derivative of $c$ along the boundary), so in our
case,
$$\{Q,{\cal O}\}= {1\over 8\pi}cc'\left(\sum_iu_i(X_i^2+4)+4a\right).
 \eqn\gomigo$$
The ghost correlation function that we need is therefore
$\langle c(\theta)cc'(\theta')\rangle$.  Using the fact that the three
ghost zero modes on the disc are $1, e^{i\theta}$, and $e^{-i\theta}$,
and normalizing the ghost measure so that $\langle c c' c''(\theta)\rangle =
1$,
we get
$$\langle c(\theta)cc'(\theta')\rangle =
2(\cos(\theta-\theta')-1).\eqn\tomigo$$
Consequently, the equation defining $S$ boils down to
$$\d S={1\over
2}\int_0^{2\pi}\d\theta\,\,\d\theta'\,\,2(\cos(\theta-\theta')-1)
\cdot
\left\langle\left({\d a\over 2\pi}+\sum_i{\d u_i X_i^2(\theta)\over 8\pi}
\right)\cdot {1\over 8\pi}\left(\sum_ju_j(X_j^2(\theta')+4)+4a\right)
        \right\rangle  .      \eqn\uppu$$
The correlation functions that arise here can all be evaluated, using
$$\eqalign{\int_0^{2\pi}\d\theta \langle X_i^2(\theta)\rangle &
=-8\pi{\partial Z \over\partial u_i}\cr
\int_0^{2\pi}\d\theta\,\,\d\theta' \langle X_i^2(\theta)X_j^2(\theta')\rangle
& = (8\pi)^2{\partial^2 Z\over\partial u_i\partial u_j},\cr}\eqn\tuppu$$
and also \longeq.
The result can be written
$$\d S=\d\left(\sum_iu_i Z-\sum_ju_j{\partial\over\partial u_j}Z+(1+a)Z\right).
     \eqn\copo$$
This can be solved by
$$S=\left(\sum_iu_i-\sum_ju_j{\partial\over\partial u_j}+(1+a)\right)Z.
       \eqn\ucop$$
Alternatively, using the fact that the $a$ dependence of $Z$ is an overall
factor of $e^{-a}$, this can be written
$$S=\left(-\sum_ju_j{\partial\over\partial u_j}-(a+\sum_ju_j){\partial\over
\partial a}     + 1\right) Z. \eqn\gogo$$
This is our final result for the space-time action for this particular
family of boundary interactions.

\chapter{Relation Between The Action And The Partition Function}

Because of the way that world-sheet path integrals can be used to compute
string interactions, some physicists have suspected (for example, see
[\fradkin,\nappi]) that the space-time
action $S$ in string theory might simply equal the world-sheet partition
function on a disc (for open strings) or a sphere (for closed strings).
Actually, since ghost zero modes make the usual partition function vanish,
the idea is really that, as long
as matter and ghosts are decoupled, $S$ would equal the matter partition
function, which I will call
$Z$.  No one has ever proposed a generalization of this conjecture
that makes sense when matter and ghosts are not decoupled.

A little reflection shows that the conjecture is more plausible on-shell
than off-shell.  The $S$ function is supposed to be gauge invariant, and
on-shell $Z$ possesses a well-known gauge invariance: it is invariant under
adding to the world-sheet Lagrangian terms of the form $\{Q,\alpha\}$
(where $\alpha$ must obey the severe restriction
that $\{Q,\alpha\}$ is independent of the ghosts,
as $Z$ is the matter partition function only).  There has never been
any indication of an off-shell gauge invariance of $Z$.

If we look back to the final result \ucop\ or \gogo\ of the last subsection,
it is clear that (in this particular approach to background independent
string theory) it is not true in general that $S=Z$.  However, these
two functions are certainly closely related.
In fact, using the formula $Z=\exp(-a)\prod_iZ_1(u_i)$, one finds
from \ucop\ that
$${\partial S\over\partial a} = Z-S. \eqn\judo$$
Consequently, on-shell or in general as long as the $a$ equation of motion
is obeyed, $Z=S$.
The purpose of the present section is to show that on-shell, $Z=S$ in general,
and to get some information about the relation between $Z$ and $S$ off-shell.

Let ${\cal V}$ be a general boundary interaction constructed from matter
fields, and take ${\cal O}=c{\cal V}$.  Let $c^{(n)}=\d^n c/\d
\theta^n$, the $n^{th}$ derivative of $c$ along the boundary.
In general, $\{Q,{\cal O}\}=\sum_{n=1}^\infty cc^{(n)}F_n$,
where $F_n$ are some matter operators.  (This follows from the fact that
$\{Q,c\}=cc^{(1)}$, while $\{Q,{\cal V}\}$, being an operator of ghost
number one without antighosts, can be expanded as a sum of expressions of
the type $c^{(n)} F_n$.)  Consequently,
in evaluating \kulko, the ghost correlation
functions that we need are of the form $\langle c(\theta)cc^{(n)}(\theta')
\rangle$ for various $n$.  In view of the form of the ghost zero
modes on the disc, these correlation functions are all linear
combinations of 1, $\cos(\theta-\theta')$, and $\sin(\theta-\theta')$.
\kulko\ can hence be written
$$\d S =\int_0^{2\pi}\d\theta\,\,\d\theta' \left\langle\d{\cal V}(\theta)
\cdot \left(A(\theta')+\cos(\theta-\theta')B(\theta')+\sin(\theta-\theta')
C(\theta')\right)\right\rangle, \eqn\huccoz$$
with $A$, $B$, and $C$ being suitable linear combinations of the $F_n$.

To proceed further, we will use the following trick (which will enable
us to avoid to have to explicitly find the generalization of the key formula
\longeq).  Suppose that the matter system consists of two decoupled subsystems.
Let $Z_1$ and $Z_2$ be the partition functions of the two subsystems,
so the combined matter partition function is $Z=Z_1Z_2$.  Let
$O_i$ be a basis of local operators for the first system, and let $\widetilde
O_j$ be an analogous basis for the second system.  The boundary interaction is
then ${\cal V}=\sum_ix^iO_i +\sum_j y^j\widetilde O_j$, with $x^i$ and $y^j$
being coupling constants of the first and second systems, respectively.
So
$$\d{\cal V}=\sum_i\d x^iO_i +\sum_j\d y^j\widetilde O_j.\eqn\mcco$$
We can expand $A$ in terms of the $O$'s and $\widetilde O$'s with
some unknown coefficients:
$$A=-\sum_i V_{(1)}^i(x)O_i -\sum_jV_{(2)}^j(y)\widetilde O_j.\eqn\rufal$$
$B$ and $C$ can be similarly expanded.
\huccoz\ can therefore be written out in terms of two point functions
of $O$'s and $\widetilde O$'s.
The terms in \huccoz\ involving $\langle O_i\widetilde O_j\rangle$ can
be very simply evaluated to give
$$ \d Z_1\cdot V_{(2)}^k{\partial\over\partial y^k}Z_2+V_{(1)}^i{\partial
\over\partial x^i}Z_1\cdot \d Z_2. \eqn\omigoo$$
The $\langle O_iO_j\rangle$ terms
contribute a one-form of the general form
$$\sum_i \d x^i a_i(x) \cdot Z_2 \eqn\muddo$$
and the $\langle\widetilde O_i\widetilde O_j\rangle$ terms contribute
a one-form that can be written
$$ Z_1\cdot \sum_j \d y^j\widetilde a_j(y),\eqn\uddo$$
with unknown functions $a_i$ and $\widetilde a_j$.   So
$$\d S=\d Z_1\cdot V_{(2)}^k{\partial Z_2\over\partial y^k}
+V_{(1)}^i{\partial Z_1\over\partial x^i}\cdot \d Z_2
+\sum_i\d x^ia_i\cdot Z_2+Z_1\cdot \sum_j\d y^j\widetilde a_j. \eqn\turboc$$

These functions can be related using the fact that $\d^2S=0$.
Calculating the exterior derivative of the one-form on the right hand side
of \turboc, and setting to zero the coefficient of
$\d x^i \d y^j$ gives
$$0=\d Z_1\left(\d y^i\widetilde a_i-\d\left(V^k{\partial Z_2\over\partial y^k}
\right)
\right)+\left( \d\left(
V_{(1)}^i{\partial Z_1\over\partial x^i}\right) -\d x^ia_i\right)
\cdot \d Z_2. \eqn\goggo$$
Using the fact that some of the quantitites entering here depend only on the
$x$'s and some depend only on the $y$'s, this implies that
$$\eqalign{\d\left(V_{(1)}^i{\partial Z_1\over\partial x^i}
\right)-a_i\d x^i&=- g\,\,\d Z_1 \cr
           \d y^i\widetilde a_i     -\d\left(V_{(2)}^j{\partial Z_2
\over\partial y^j}\right)
            & = g \,\,\d Z_2\cr} \eqn\nuucu$$
for some constant $g$.
Using these formulas to express the $a_i$ and $\widetilde a_j$ in terms of
the $Z$'s and $V$'s, \turboc\ can be rewritten
$$\d S=\d\left(\left(V_{(1)}^i{\partial\over\partial x^i}+V_{(2)}^j{\partial
\over\partial y^j} + g\right)Z_1Z_2\right).     \eqn\muucu$$
And so
$$ S=\left(V_{(1)}^i{\partial\over\partial x^i}+V_{(2)}^j{\partial
\over\partial y^j} + g\right)Z_1Z_2.     \eqn\mmuucu$$

So far, we have considered ``matter'' to consist of two decoupled
systems, but this restriction is unnecessary.  To any matter system
of interest one can always add an auxiliary decoupled system
and carry out the above analysis.  Then the auxiliary system
can be suppressed by setting its couplings to a fixed value.
The conclusion is that in general there is a vector field $V$ on the space
of world-sheet theories, and a constant $g$, such that
$$S=\left(V^k{\partial\over\partial x^k}+g\right) Z. \eqn\mcmcmmc$$
Moreover, $V$ vanishes for classical solutions, since it was constructed
from $A$ and so ultimately from $\{Q,{\cal O}\}$.
Therefore, on-shell $S$ is a constant multiple of $Z$, as promised.
Our earlier formula \gogo\ serves to illustrate
\mcmcmmc\ for some
particular boundary interactions.
\refout
\end

\end